\documentclass[useAMS,usenatbib]{mn2e}

\usepackage{graphicx}
\usepackage{caption}
\usepackage{subcaption}
\usepackage{epstopdf}
\usepackage{amsmath}
\usepackage{lmodern}
\usepackage{slantsc}

\def\gs{\mathrel{\raise0.35ex\hbox{$\scriptstyle >$}\kern-0.6em
\lower0.40ex\hbox{{$\scriptstyle \sim$}}}}
\def\ls{\mathrel{\raise0.35ex\hbox{$\scriptstyle <$}\kern-0.6em
\lower0.40ex\hbox{{$\scriptstyle \sim$}}}}
\def\ls{\mathrel{\hbox{\rlap{\hbox{\lower4pt\hbox{$\sim$}}}\hbox{$<$}}}}
\def\gs{\mathrel{\hbox{\rlap{\hbox{\lower4pt\hbox{$\sim$}}}\hbox{$>$}}}}

\newcommand\gtsim{\mathrel{\lower0.6ex\hbox{$\buildrel {\textstyle >}
  \over {\scriptstyle \sim}$}}}
\newcommand\ltsim{\mathrel{\lower0.6ex\hbox{$\buildrel {\textstyle <}
  \over {\scriptstyle \sim}$}}}
\newcommand\etal{\mbox{{et al.\ }}}
\newcommand{\degree}{$^{\circ}$ }

\title[]
      {Environments and Morphologies of Red Sequence Galaxies with Residual Star Formation in Massive Clusters}
\author[Crossett, Pimbblet, Stott, Jones]
       {Jacob P.\ Crossett$^{1,2}$\thanks{email: jake.crossett@monash.edu}, Kevin A.\ Pimbblet$^{1,2,3}$, John P.\ Stott$^{4}$, D. Heath Jones$^{1,2}$ 
        \vspace*{1mm}\\
        $^{1}$School of Physics, Monash University, Clayton, Victoria 3800, Australia\\
        $^{2}$Monash Centre for Astrophysics (MoCA), Monash University, Clayton, Victoria 3800, Australia\\
        $^{3}$Department of Physics, University of Oxford, Denys Wilkinson Building, Keble Road, Oxford OX1 3RH, UK\\
        $^{4}$Department of Physics, University of Durham, South Road, Durham, DH1 3LE\\}

\pagerange{000--000}

\begin{document}

\maketitle

\begin{abstract}
We present a photometric investigation into recent star formation in galaxy clusters at $z\sim0.1$. We use spectral energy distribution templates to quantify recent star formation in large X-ray selected clusters from the LARCS survey using matched \textsl{\textsc{galex}}
NUV photometry. These clusters all  have signs of red sequence galaxy recent star formation (as indicated by blue NUV-R colour), regardless of cluster morphology and size. A trend in environment is found for these galaxies, such that they prefer to occupy low density, high cluster radius environments. 
The morphology of these UV bright galaxies suggests that they are in fact red spirals, which we confirm with light curves and Galaxy Zoo voting percentages as morphological proxies. These UV bright galaxies are therefore seen to be either truncated spiral galaxies, caught by ram pressure in falling into the cluster, or high mass spirals, with the photometry dominated by the older stellar population.
\end{abstract}

\begin{keywords}
galaxies: elliptical and lenticular, cD - galaxies: evolution - galaxies: stellar content - galaxies: clusters: general - ultraviolet: galaxies 
\end{keywords}

\section{Introduction}
The stellar populations of galaxies are reflected in their colours and magnitudes. Morphologically elliptical and lenticular (henceforth called early-type) galaxies are generally passively evolving red and dead galaxies, with optical colours tightly constrained to a red sequence (Faber 1973; Visvanathan \& Sandage 1977). This red sequence is seen to be a universal property of clusters, showing a tight relation with a small scatter due to metallicity and age effects (Bower et al.\ 1992; Kodama et al.\ 1998). Morphologically spiral (or late-type) galaxies, however, are blue, star forming and are generally found in a large scattered blue cloud below the red sequence relation on the colour magnitude diagram. While there are galaxy populations of optically red spirals (e.g. Wolf \etal 2009; Masters \etal 2010), and blue ellipticals (Schawinski \etal 2009; Huertas-Company \etal 2010; McIntosh \etal 2013), the optically-defined red sequence will recover the majority of morphologically elliptical (early-type) galaxies in a cluster.

The colour-magnitude relationship is less tight if one looks in the NUV ($\sim2250$ \AA), with high amounts of scatter in the NUV-r and NUV-J colour magnitude relations for red sequence galaxies (Yi \etal 2005; Kaviraj \etal 2007; Rawle \etal 2008). There are many galaxies within this UV colour magnitude relation, that show a ``high UV excess", despite passive optical colours (MacLaren \etal 1988). These UV ``blue" galaxies are believed to show signs of recent ($\sim$1 Gyr) low level ($\sim 1\%$ of the stellar mass) star formation (Yi \etal 2005; see also Schawinski et al.\ 2007; Salim et al.\ 2012; McIntosh et al.\ 2013). Moreover, this star-formation appears exclusively in fast rotating systems as noted by the \textsc{Sauron} survey (Shapiro et al.\ 2010).

Recent star formation has been suggested as observational evidence for a ``frosted" population of young stars in stellar population modelling (Trager \etal 2000). These models predict a second burst of young stars, forming much later than the large initial parent population, improving the agreement of these models to observations (Trager \etal 2000). These two burst models are best described by the presence of a burst 1 Gyr ago, with mass fractions between 1 and 10 \% (Ferreras \& Silk 2000; Serra \& Trager 2006). The frosted models also significantly improve the existing stellar population models by explaining the observed excess UV flux (Rawle \etal 2008).

The frosted model of star formation, with high UV flux from young stars is believed to occur following a recent minor or major merger. Many of the UV bright galaxies have been known to have disturbed, and even spiral morphologies, compared to the their passive red galaxy counterparts (Yi \etal 2005; Kaviraj \etal 2011). Frosted models have been used in modelling the stellar populations of mergers, with small amounts of young stars supporting the argument of mostly ``dry", gasless mergers, compared to a star-formation inducing ``wet'', gas rich merger (Sanchez-Blasquez \etal 2009). Little evidence has been seen for other mechanisms to cause this recent star formation, with no trend in density or cluster environment (Schawinski \etal 2007; Yi \etal 2011).

Alternatively, these ``recent star formation" galaxies could be an example of a disk galaxy that has been recently truncated by an external source. This truncation of star formation can change the optical colours of a galaxy, but leave the morphology intact (Masters et al.\ 2010). These galaxies are generally large spirals ($log_{10}M\sim10^{10}$), and lower star formation rates compared to blue spirals of the same mass (Masters et al.\ 2010; Wolf et al.\ 2009). These galaxies show significant differences in star formation within 500 Myr, but maintain similar amounts of recent star formation, suggesting these galaxies to have been truncated blue spirals (Tojeiro et al.\ 2013). Other evidence suggests that these red spirals are still forming stars, but are optically red due to their mass -- the specific star formation rate has decreased without a truncation (Cortese 2013). In this scenario, the UV will still probe the recent star formation at a residual level, as these galaxies are simply dominated by their large older stellar population due to their large mass (Cortese 2013). This population may harbour high UV flux, as a possible cause of UV scatter on the colour magnitude plane.

One major alternative for this UV excess is older stars with a UV upturn effect. This upturn in UV is currently believed to be caused by high metallicity low envelope mass, horizontal branch stars (Code \& Welch 1979; Burstein \etal 1988; Greggio \& Renzini 1990; see O'Connell 1999 for a review). This UV upturn process is thought to be caused by extra helium within these stars, forcing faster burning of the hydrogen envelope, causing extra mass loss and increasing the rest frame FUV flux (Yi \etal 1997). Models of the UV upturn helium enrichment include effects with helium sedimentation within the galaxies parent cluster, where the intracluster medium is not homogeneously dispersed, but with helium and metals falling to the centre of the gravitational well (Peng \& Nagai 2009). Other scenarios involve various binary interactions and the creation of hot subdwarfs that can reproduce the effects of the UV upturn (Han \etal 2007).
Interference from the UV upturn from the rest frame 1550\AA \ into the NUV, a ``leaking" of UV upturn flux, has been explored as an alternative origin of these UV bright sources (Smith \etal 2011). This can blur the difference between the residual star formation within a galaxy, and older population contributions. While these do not impair results at earlier epochs (Kaviraj \etal 2011), care needs to be taken to define limits of older stellar population contamination in low redshift ellipticals (Kaviraj \etal 2007).

In this study we aim to determine and explore the origin of UV bright galaxies in large galaxy cluster environments, separating the role of young and old stellar populations in UV flux, as well as determining the mechanism for blue UV colour in red sequence galaxies. Galaxy clusters are an ideal laboratory for studying the effects of galaxy interactions, with many forms of interactions (both between galaxies, and with the cluster medium) able to change galaxies stellar populations and morphology. While previous studies have suggested the presence of widespread recent star formation in clusters (e.g. Rawle et al.\ 2008), or have investigated the density and radial profiles of recent star formation (Schawinski et al.\ 2007; Yi et al.\ 2011), this is the first study to investigate the effect of cluster environment for these NUV bright sources. In this work, we aim to determine the mechanisms and conditions suitable for either recent star formation, or UV upturn stars, from UV bright sources. To do this, we use several large X-ray selected clusters from the optical LARCS dataset (Pimbblet et al.\ 2001; 2002; 2006), with known red sequence information, in conjunction with \textsl{\textsc{Galex}}
 NUV data (Martin et al.\ 2005).

In Section~2 we outline the data and reduction processes used, including photometric background subtraction and red sequence fitting. Section~3 details our definitions of ``UV Bright" galaxies, and derives UV bright fractions for each individual cluster and for a composite sample. We further investigate the properties of these UV bright red sequence galaxies in Section~4, focussing on the role of density and cluster radius as a metric for galaxy environment. The physical nature of these galaxies is investigated in Section~5, looking at the morphologies and light curves of the UV bright population. We discuss potential mechanisms and causes for these UV bright galaxies in Section~6, before summarising our main findings in Section~7. Throughout this work, we assume a standard cosmology with values of $H_0$=71 km/s/Mpc and $\Omega_v$=0.23.

\section{Sample}
\subsection{Las Campanas/AAT Rich Cluster Survey}
The Las Campanas/AAT Rich Cluster Survey (LARCS; Pimbblet et al.\ 2001) consists of panoramic ($\sim 2$ degree diameter fields; roughly $\sim10$ Mpc radii at typical cluster redshift) high quality spatial resolution in $B$- and $R$-bands (sourced from the 1m Swope Telescope at Las Campanas Observatory) of X-ray selected bright ($L_X>5 \times 10^44$ erg s$^{-1}$) galaxy clusters, thus largely eliminating aperture bias.  Pimbblet et al.\ (2001) gives the detail of the observations, reduction and analysis of this optical imaging and we summarise the global parameters of the clusters used in this work in Table~\ref{tab:clusters}. Here, we note that the ($B-R$) colours derived from these images  use 4'' apertures which corresponds to $\sim10$ kpc at the typical redshift of these galaxy clusters. The LARCS photometric zeropoints are tied to Landolt (1992) standards and the final photometric accuracy is better than 0.03 mags and the catalogues are at least 80 per cent complete at a depth of $B\sim23$ and $R\sim22$. The galaxy-star separation techniques are analogous to Pimbblet \etal (2001), taking into account a full width half maximum (FWHM) to cut out compact sources, in conjunction with \textsc{SExtractor}'s in built neural network based star classifier, \textsc{Class\_star}, or P(*). Taking the FWHM$>2$ and P(*)$<0.5$, we can eliminate most stellar sources, with a misclassification rate of $<3\%$ (Pimbblet \etal 2001).

\subsection{\textsl{\textsc{GALEX}}}
We complement the optical data with UV data from \textsl{\textsc{Galex}}
 General Release 6, taking tiles of matching position to each cluster from the all sky survey (AIS) as well as limited overlapping regions from the medium imagine survey (MIS) (Martin et al.\ 2005). The \textsl{\textsc{Galex}}
 satellite provides high quality 1.2\degree field NUV imaging (1750-2750\AA) down to a magnitude of 20.5 (AB) in the AIS, and 23.5 in the MIS (Martin \etal 2005; Figure 1). These images are resolved with a full width half maximum of 5.6 arcseconds. The use of two different exposure times of the two surveys (0.1ks for the AIS and 1.5ks for the MIS; Martin \etal 2005) can potentially cause an exposure bias in finding NUV bright galaxies. To limit any bias between clusters, we force a limit on the MIS data at an upper bound magnitude of 23, to match the limiting magnitude in our sample of AIS objects. 

In the subsequent work, we utilise the full field of view of \textsl{\textsc{Galex}} to work with. However, Morrissey et al.\ (2007) notes that the outer $\sim 5$ per cent of the fields have both poorer astrometric solutions and NUV photometric zeropoints compared to the central region (cf.\ Drinkwater et al.\ 2010). Excluding detections at these large radii only affects 3 sources, and does not affect the results presented in the subsequent work. Taking a more conservative 1.15 degree field of view, we eliminate the outermost 16 sources from our sample. This also has no significant effect in both the cluster radius and density relations discussed below.

\begin{figure}
\begin{center}
\includegraphics[scale=0.35]{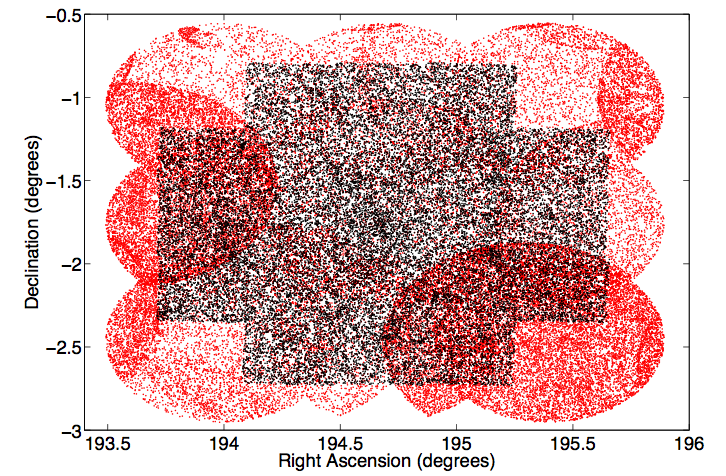}
\caption{Sources in the region of Abell 1650 from \textsl{\textsc{Galex}}
 NUV (red), and LARCS optical (black). The different survey density densities are apparent in the \textsl{\textsc{Galex}}
 imaging, with varied depths. This was corrected by introducing a limiting magnitude in the NUV.}
\label{fig:areas}
\end{center}
\end{figure}

\subsection{Early-type Selection and Background Correction}
In order to determine the environment of these UV bright cluster galaxies, it is imperative to eliminate foreground/background interlopers from the cluster. In the absence of velocity measurements, this process becomes non-trivial. Pimbblet et al. (2002) detail a robust statistical method of background subtraction that has been repeated throughout the literature for different applications (Urquhart et al. 2010; Valentinuzzi et al. 2011; Pimbblet \& Couch 2012). Here, we summarise the key details of this correction:

In this method, galaxy sources are taken from a region outside each cluster, where sources are labelled as `field' In this study, the large panoramic view from the LARCS images allow the outer edges of each cluster sample to be used as `field' points. Galaxy sources within each uncorrected cluster (denoted cluster+field) are binned in a 10 x 10 grid overlaid on a B-R (optical) colour magnitude relation, and similarly for the combined outer regions of the LARCS images (denoted field). From this the probabilities of galaxies being cluster members are calculated via the equation:
\begin{equation}
P(Cluster)=\frac{N(Cluster+Field)-A \times N(Field)}{N(Cluster+Field)}
\end{equation}
where A is a scaling value to account for the difference in area between the sources radius and the background. Using randomly generated numbers, each individual galaxy is then included or excluded based on this probability, and repeated over 100 trials in a Monte Carlo fashion, to generate a field corrected sample.

The difficulty of this process arises when a binned section of the background is larger than the corresponding cluster bin, creating a negative probability for a cluster galaxy. To avoid this problem, an adaptive bin size is employed, whereby any square with a 'negative galaxy' problem has the bin size increased to include the adjacent bins. This makes a 2 x 2 size bin from which deduction is re-computed, without the negative square.

We use definitions from Pimbblet et al.\ (2002) to determine the red sequence , where a biweight fit was used to suggest the appropriate locus on the optical colour magnitude diagram from the statistically corrected sample in each cluster. The $1\sigma$ uncertainties arise from uncertainties of 100 trials of the statistical correction described above. 
We note a caveat in this method of determining the red sequence, where the fit used from (Pimbblet et al.\ 2002) is derived from the red sequence fit of the inner 1 Mpc of each cluster. This study found the red sequence becomes progressively more blue with increasing radius, in line with the morphology density relationship (Dressler \etal 1980). This method, therefore, may not encompass only the morphological early-type galaxies, with a fluctuation of the red sequence in each cluster.
We note, however, that this fit becomes a conservative measure of the most optically red objects in these clusters, even to high radii. While we do not allow for more optically blue elliptical galaxies further out in the cluster, we force objects are the most optically passive to be used throughout. This makes our sample less likely to have interference from spirals at higher radii. 
We also note the necessity of using a previous definition for a red sequence in preventing any bias from a UV limited sample. Re-defining a red sequence using similar methods to  make a colour magnitude relation of a UV luminous sample may in fact cause the baseline of the sample to favour optically blue objects.

\begin{table*}
\begin{center}
\caption{The sample of clusters used in this work.  
The velocity data is sourced from Pimbblet et al.\ (2006) The coordinates specify the X-ray centre of the cluster. The X-Ray luminosities ($L_X$) are sourced from the Base de Don\'{e}es Amas de Galaxies X (BAX; webast.ast.obs-mip.fr/bax; Sadat et al.\ 2004). The virial radius ($R_{virial}$) is computed following Carlberg et al.\ (1997) via $R_{vir} \sim r_{200} \sim 
\frac{\sqrt{3} \sigma_z}{10 H(z)}$ where $H(z)^2 = H_0^2 (1+z)^2 (1+\Omega_0 z)$.
\hfil}
\begin{tabular}{lllllll}
\noalign{\medskip}
\hline
Name  & RA      & Dec     & $L_X$                           & $\overline{cz}$ & $\sigma_{cz}$  & $R_{virial}$  \\
      & (J2000) & (J2000) & ($\times 10^{44}$ erg s$^{-1}$) & (km\,s$^{-1}$)  & (km\,s$^{-1}$) & (Mpc)      \\
\hline
Abell~22 & 00:20:38.64 & -25:43:19 & 5.31 & $42676 \pm 98$ & $806^{+80}_{-62}$  & 2.3 \\ 
Abell~1084 & 10:44:30.72 & -07:05:02 & 7.06 & $39762 \pm 64$ & $712^{+50}_{-42}$  & 2.0 \\
Abell~1650 & 12:58:41.76 & -1:45:22 & 7.81 & $25134 \pm 55$ & $795^{+42}_{-36}$  & 2.4 \\
Abell~1664 & 13:03:44.16 & -24:15:22 & 5.36 & $38468 \pm 96$ & $1069^{+75}_{-62}$  & 3.1 \\
Abell~2055 & 15:18:41.28 & 06:12:40 & 4.78 & $30568 \pm 101$ & $1046^{+80}_{-65}$  & 3.1 \\
\hline
\noalign{\smallskip} 
\end{tabular}
  \label{tab:clusters}
\end{center}
\end{table*}

\section{UV Bright Galaxies}
In this section we use NUV-R colour magnitude diagrams to investigate the presence of UV bright red sequence galaxies. We analyse the validity of thresholds for UV bright galaxies from previous works, and use methods to determine the amount of UV bright galaxies in each cluster. The colour magnitude diagrams use the NUV-R colour, for best overlap with previous works (Kaviraj \etal 2007). 

\subsection{SED fitting}
\begin{table}
\caption{Synthetic colours from spectral templates from Brown et al.\ (2013, in prep). Galaxies with older stellar populations fall above a value of NUV-r=5.4 for a redshift of 0.1, strengthening the use of 5.4 as a recent star formation threshold (Kaviraj \etal 2007).}
\begin{center}
\begin{tabular}{lclclc|c|c|}
\noalign{\smallskip}
\hline
\#NGC&Class& $FUV-NUV$ & $NUV-R$ & $Abs(R)$\\
\hline
0584 &Passive & 0.82& 6.57&-21.50 \\
4125 &Passive & 1.07& 6.53&-22.94 \\
4552 &UV-upturn & 0.29& 5.89&-22.59 \\
4889 &UV-upturn & 0.38& 5.93&-24.37 \\
5195 &Post starburst & 0.57& 5.39&-21.09 \\
7331 &Post starburst& 0.87& 4.85&-23.09 \\
\hline
\noalign{\smallskip} 
\end{tabular}
\end{center}
\label{tab:standards}
\end{table}

This study will look at the NUV colours of ``UV bright" galaxies in clusters. However, the threshold for UV bright, is not formally defined. Previous studies have used flux limits (Yi et al.\ 2005), and colour cuts based on UV upturn samples (Kaviraj et al.\ 2007; Jeong et al.\ 2009). Previous studies have used a reference UV upturn galaxy, NGC 4552, reported as one of the strongest known UV upturn galaxies as a reference threshold (Kaviraj et al.\ 2007; Schawinski et al\ 2007; Jeong et al.\ 2009).

\begin{figure*}
\begin{center}
\includegraphics[scale=0.45]{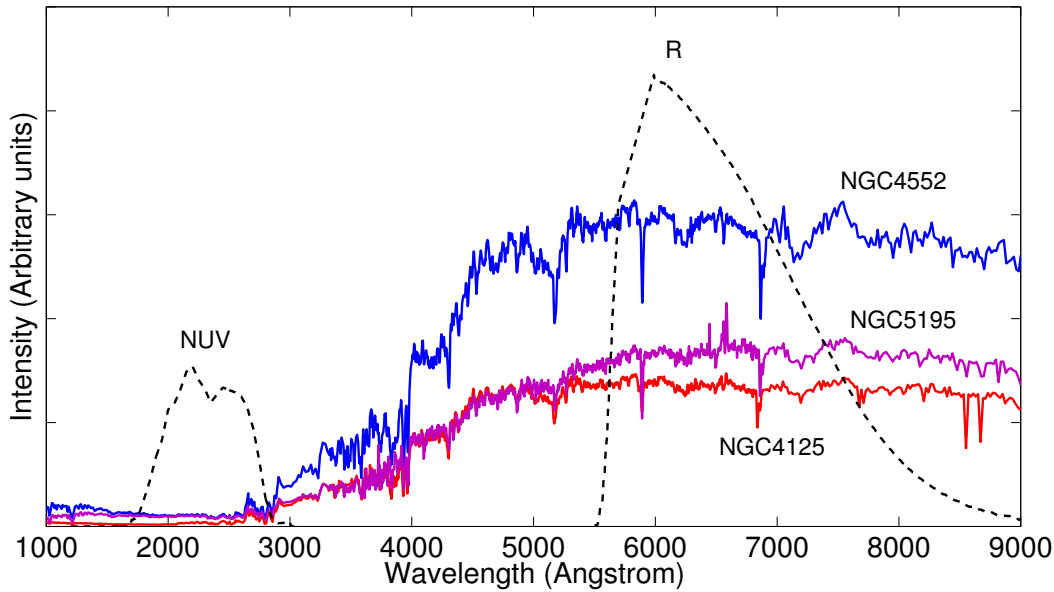}
\caption{Relative spectra  of three template galaxies, NGC 4125 (red), NGC 4552 (blue), and NGC 5195 (magenta). The spectra of NGC5195, the post starburst galaxy, closely matches that of the passive elliptical (NGC4125) in the range 3000-7000 \AA, but then changes to match the UV upturn galaxy NGC 4552 at $<2500$ \AA. The filters for NUV and R (dotted lines respectively) denote the relative positions and strengths of the filters used.}
\label{fig:spectra}
\end{center}
\end{figure*}

In order to accurately separate populations of UV bright galaxies, we test existing definitions of these studies and include several reference galaxies for a UV threshold. We employ spectral energy distribution (SED) templates from Brown et al.\ (2013; in prep) to test the UV properties of idealised galaxies types. By creating synthetic colours in the the UV and R bands, we can compare the NUV-R colours to previously defined thresholds and determine the most suitable. Filter transmission curves in the FUV, NUV and R bands are taken and the flux of each galaxies spectra is integrated over these curves. This allowed an ideal flux in each hypothetical filter allowing the FUV-NUV and NUV-R colours to be calculated (Figure \ref{fig:spectra}). Using these colours, the UV properties of these known galaxies formed a diagnostic, to test the thresholds of quiescent, UV excess, and post starburst populations UV colours.

Six spectral energy distribution templates are used to test the validity of the UV threshold (Table \ref{tab:standards}). These included two previously known post starburst galaxies (NGC 5195 and 7331), two elliptical galaxies (NGC 0584 and 4125), and two UV upturn galaxies (NGC 4552 and NGC 4889). While the post starburst galaxies are known from previous literature, the UV upturn SEDs are chosen based on a FUV-NUV criterion from Yi \etal (2011), where the FUV-NUV colour $<0.9$ is considered to be significantly blue to be UV upturn. 

\begin{figure*}
        \centering
        \begin{subfigure}{1\textwidth}
                \includegraphics[scale=0.45]{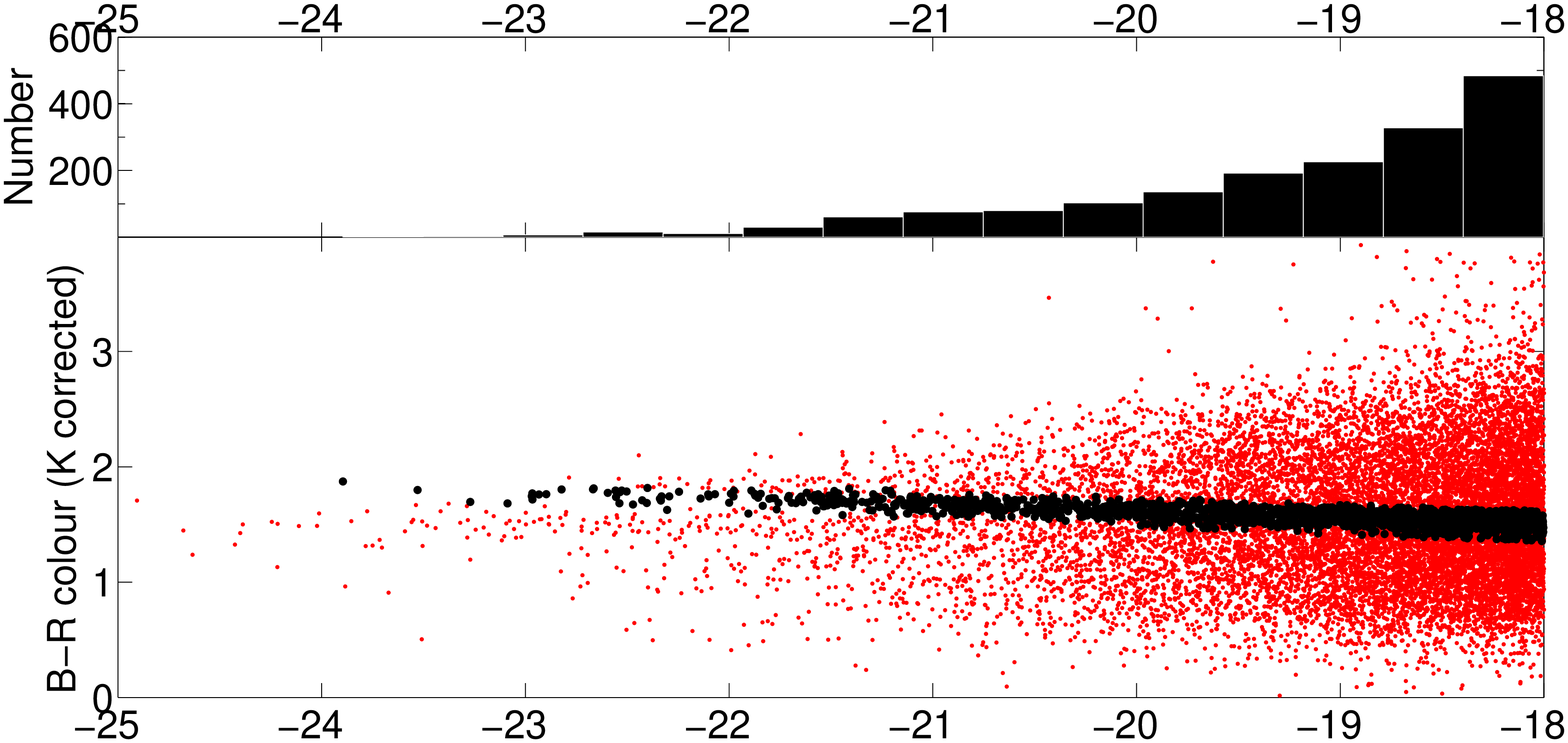}
        \end{subfigure}%
      		\vspace{4mm}
        \begin{subfigure}{1\textwidth}
                \includegraphics[scale=0.45]{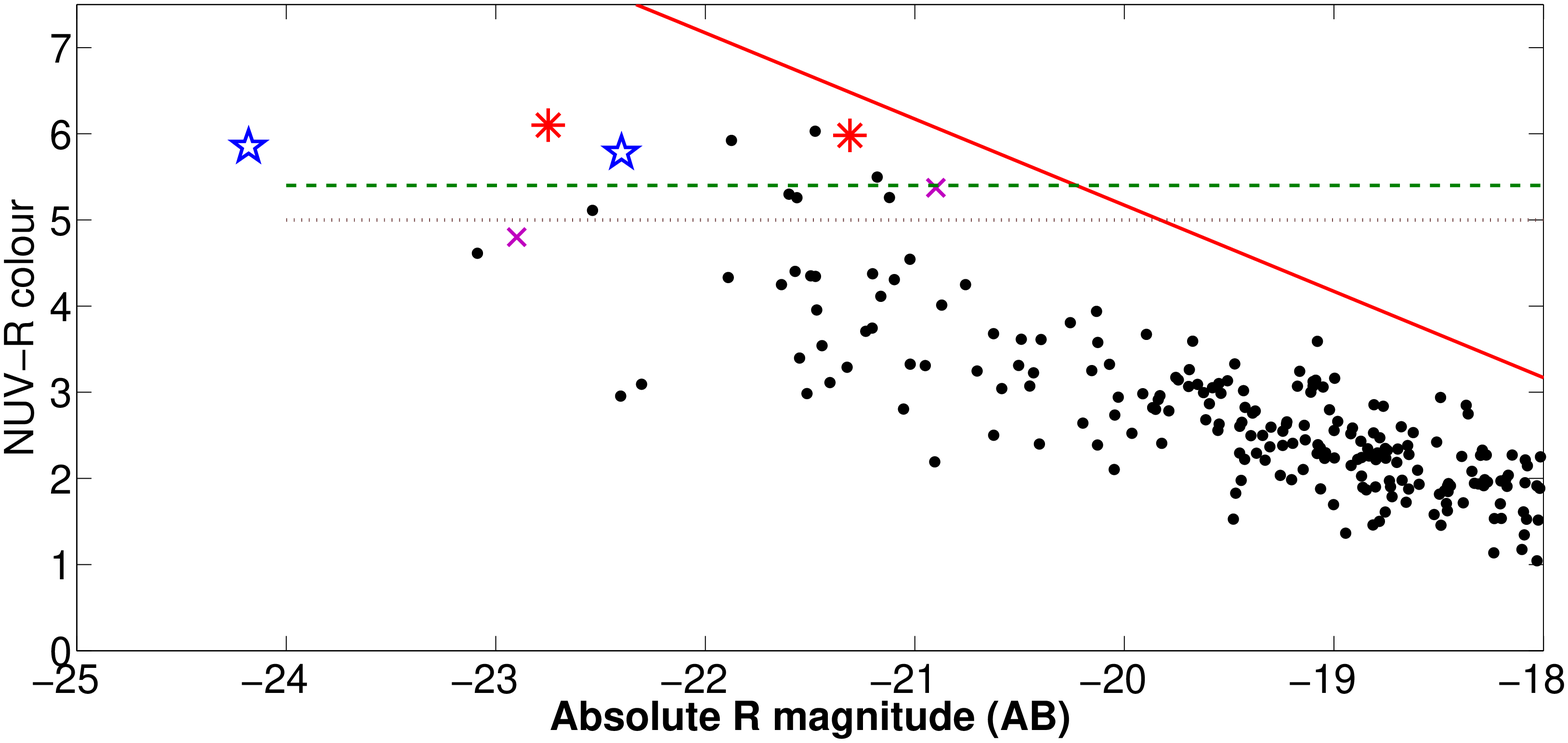}
        \end{subfigure}
       \caption{(Top) K-corrected optical colour-magnitude diagrams for the selected red sequence galaxies (black) down to a cut of R=-18. Red points are non-red sequence galaxies in the cluster area. (Bottom) The UV relation shows a high scatter in detected \textsl{\textsc{Galex}}
 elliptical galaxies, despite the hard optical colour cut for the red sequence. The green dashed line shows the threshold for recent star formation from Kaviraj \etal (2007), and the brown dotted line shows thresholds from Jeong \etal (2009). The red line gives the completeness of the all sky data from Galex for our data lmits. Red Stars (passive eliptical galaxies) and blue stars (UV upturn galaxies) do not cross the threshold, whereas the post-starburst galaxies (pink crosses) show bluer NUV-R colours.}
\label{fig:CMRS}
\end{figure*}

We note that although our templates include fair representations of elliptical and UV upturn galaxies, the post starburst templates do not come from homogenous hubble types. NGC 5195 is a dwarf elliptical, in a current merger with NGC 5194 (Kohno et al. 2002), while NGC 7331 is known to be a spiral galaxy, with a recent starburst (Tosaki \& Shioya, 1997). While these are not recent star formation in classical ellipticals, the use of these different sources provides a benchmark for a variety of morphologies and merger histories. Figure \ref{fig:CMRS} shows the template SED colour against the NUV-R red sequence galaxies.

This shows that galaxies with older stellar populations (the quiescent ellipticals and UV upturn galaxies) do not fall below the recent star formation threshold given in Kaviraj et al.\ (2007) of NUV-R=5.4. Galaxies previously known to be in a post-starburst phase however, do appear to fall on or below the line at redshift of 0.1. The threshold of Jeong \etal (2009), of NUV-R=5.0 does not encompass both of the post starburst galaxies. To test for invariance, both definitions used to compare the radial and density profiles (see Section 4). The threshold for recent star formation is set to blueward of NUV-R=5.4 throughout, unless otherwise noted.

\subsection{Recent Star Formation in Clusters}

Using the NUV-R threshold described in section 3.1, fractions of galaxies deemed to contain high UV, or have recent star formation, are computed for 100 realisations of the statistical background correction. All galaxies with null detections in the NUV band are assumed to belong to the passively evolving galaxies. The fractions for each of the five clusters is taken, with $1\sigma$ uncertainties from the statistical background uncertainties from section 2.3. The clusters show a general trend of around 10\% of red sequence galaxies exhibiting signs of  recent star formation in most clusters. Due to the extensive exposure in the medium imaging survey for Abell 2055 the fractions of UV bright galaxies increase. This makes the fraction in line with the accepted values of recent star formation (30\%, Kaviraj \etal 2007). Comparisons between Regular (R) and Irregular (I) clusters determined by Pimbblet \etal (2002), and the relevant fractions of recent star formation (RSF) are included in Table 3, showing little difference between clusters based on their cluster morphology.

\begin{table}
\centering
\caption{Fractions of UV bright galaxies for each cluster in the sample. Most of these galaxies contain approximately 10\% recent star formation, aside from Abell 2055. While an NUV magnitude limit was imposed, the high amount of MIS imaging for Abell 2055 has increased the fraction of RSF, in line with previous \textsl{\textsc{Galex}}
 MIS studies.}
\begin{tabular}{lclclcl}
\hline
Name & Cluster type & UV bright \%\\
\hline
Abell~22 & Regular&$11\pm2$  \\
Abell~1084 &Irregular & $9\pm1$\\
Abell~1650 &Regular & $9\pm1$\\
Abell~1664 &Irregular & $11\pm1$\\
Abell~2055 & Regular & $35\pm2$\\
\hline
\end{tabular}
\label{tab:fractions}
\end{table}

\section{Environmental Variation}
\begin{figure}
        \centering
        \begin{subfigure}[b]{1\textwidth}
                \includegraphics[scale=0.3]{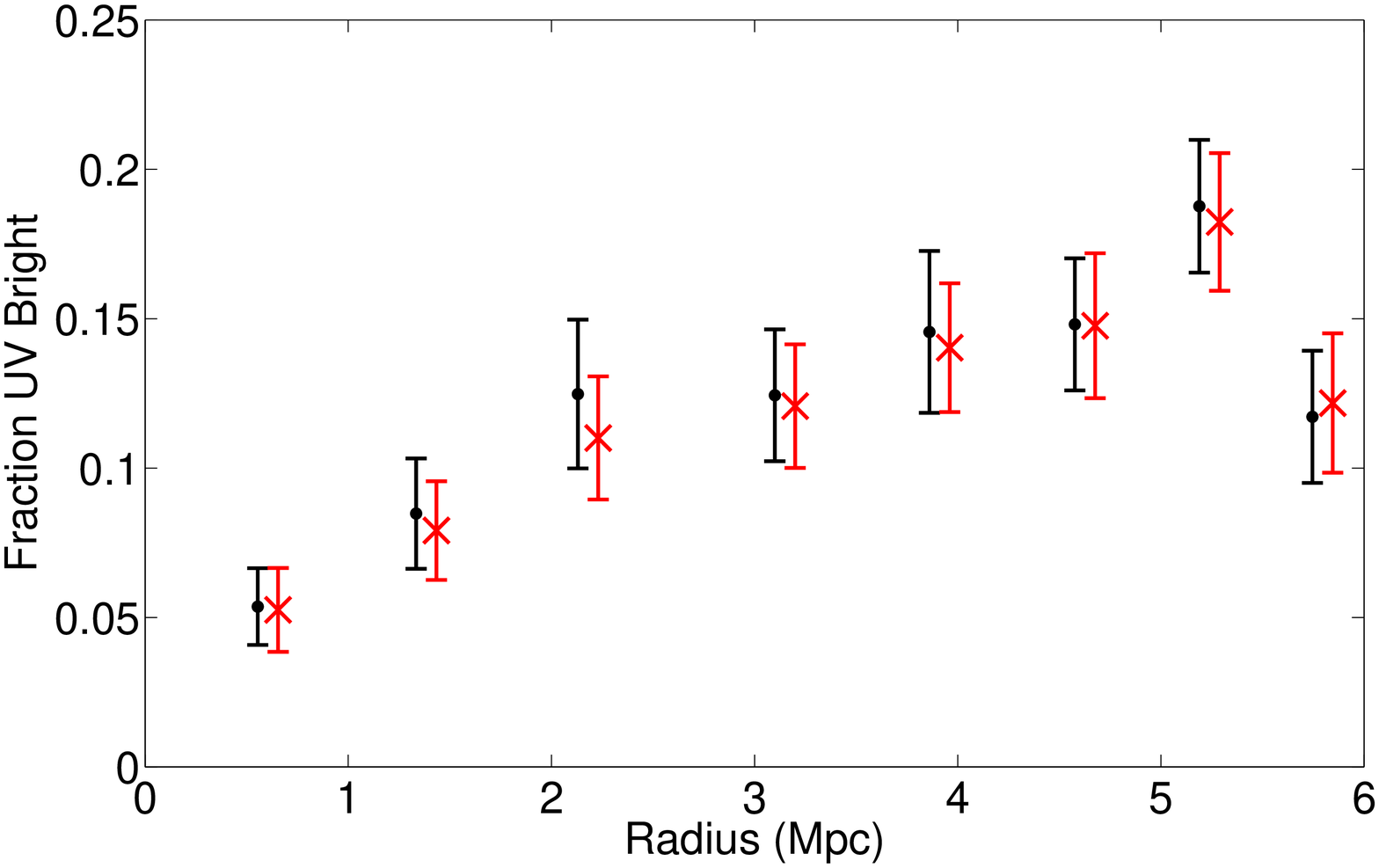}
        \end{subfigure}%
      		\vspace{4mm}
        \begin{subfigure}[b]{1\textwidth}
                \includegraphics[scale=0.3]{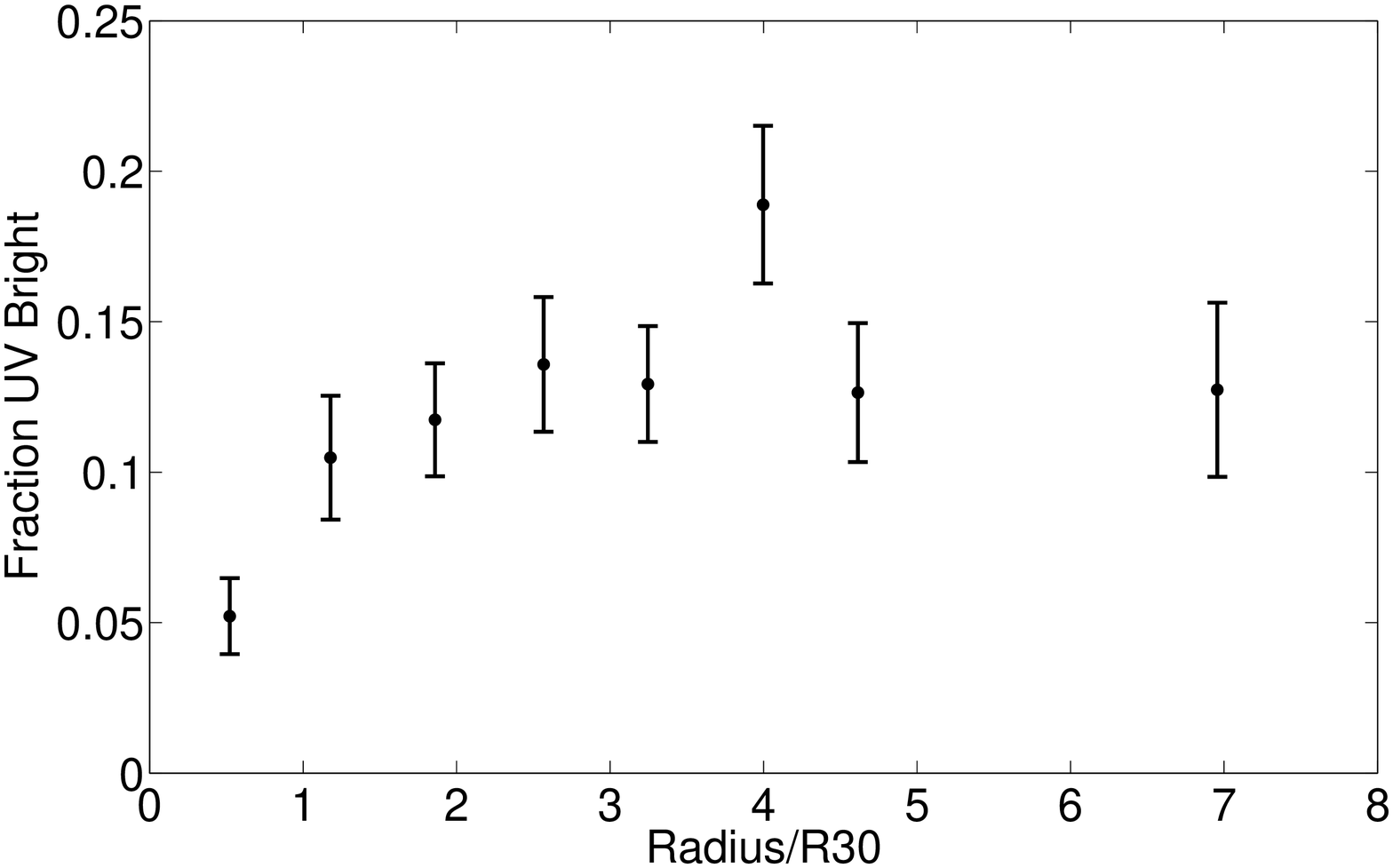}
        \end{subfigure}
       \caption{(Top) Fractions of UV bright galaxies for different cluster radius bins. The uncertainties are derived from the statistical background correction used, described in 2.2. The colours represent a comparison of two different threshold points from Kaviraj \etal (2007; at 5.4, black points) and Jeong \etal (2009; at 5.0, red crosses), showing a similar trend regardless of choice. Red crosses are offset from the original black points by an arbitrary amount for clarity. All of the bins contain equal members. (Bottom) Fractions of UV bright galaxies as a function of r30, using the threshold from Kaviraj \etal (2007). As a surface approximation, both the spacial radius and r30 radii show a dip in galaxies with RSF towards the cluster core.}
\label{fig:radfrac}
\end{figure}

\begin{figure}
\begin{center}
\includegraphics[scale=0.3]{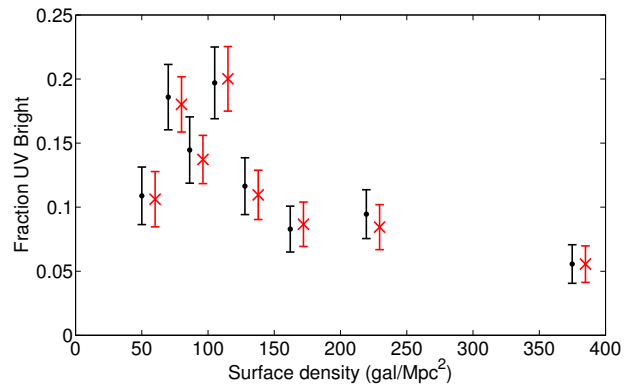}
\caption{Fractions of recent star formation for surface density binned data, with uncertainties from the statistical background correction. Colours represent two different threshold points (5.4, black points, 5.0, red crosses), showing a similar trend regardless of choice. Red crosses are offset from the original black points by an arbitrary amount for clarity. High surface densities show a deficiency of UV bright galaxies, compared to less dense regions.}
\label{fig:denfrac}
\end{center}
\end{figure}

After quantifying the amount of UV bright red sequence galaxies within galaxy clusters, we make a combined sample from all 5 clusters using a fixed cluster-centric radial distance metric. Using this combined sample, fractions of UV bright galaxies are made as a function of cluster radius as a proxy for environmental variation. In addition to this, the UV bright fraction as a function of 2D surface density, using a 10th nearest neighbour metric in the LARCS R band, is calculated. This approach is used to better isolate the local density of a galaxy, instead of its large scale density (Muldrew et al.\  2012) 
Figures \ref{fig:radfrac} and \ref{fig:denfrac} show galaxies binned by their respective environmental metric, and the fraction of galaxies with recent star formation taken for each bin over 100 background trials. The figures show a combined tendency for recent star formation to occur in the outer radii of a cluster, and in low density areas. Cluster cores, and high density regions appear to have a factor of 2 decrease in the recent star formation, compared to the higher radius counterparts.

In figures \ref{fig:radfrac} and \ref{fig:denfrac} we show a second trend (marked in red), which is the threshold of Jeong \etal (2009) to compare the different threshold of NUV-R=5.0. The trends show a small offset but no significant change in trend. The most UV bright objects preferentially lie within the low density/high cluster radii regions, regardless of the threshold used. It is assumed that UV bright galaxies are generally found in low density environments, irrespective of the definition of threshold. 

\begin{figure}
\begin{center}
\includegraphics[scale=0.27]{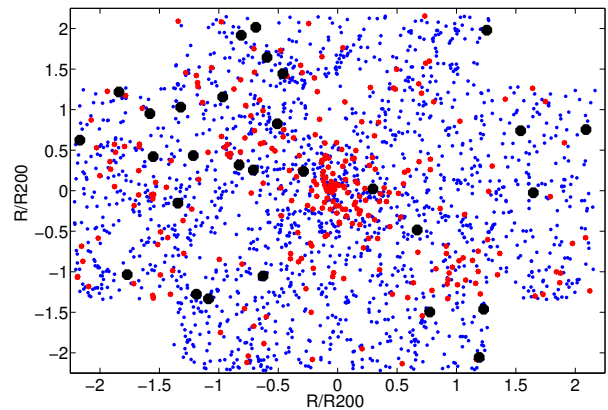}
\caption{Uncorrected galaxy locations in Abell 1650 with Red Sequence (red), UV bright (black), and other galaxies (blue) displayed as a 2d projection on the sky. The UV bright galaxies generally reside away from the cluster centres, in low density regions, despite the numbers of passive red sequence galaxies in the core regions.}
\label{fig:locations}
\end{center}
\end{figure}

Figure \ref{fig:radfrac} shows the variation of recent star formation with projected cluster radius, and $R_{30}$ scaled radius (top, and bottom respectively), while Figure \ref{fig:denfrac} shows surface density. There is a trend towards low density and high cluster radii galaxies being UV bright. Over the density and radial ranges covered, a factor of three increase is seen in the fraction of recent star formation galaxies for higher radius galaxies, and for lower density galaxies. This suggests that galaxies with recent star formation are likely to be found on the edges of the cluster, and have not evolved for long times within the cluster itself. Figure \ref{fig:locations} shows the positions of the UV strong galaxies on a 2-D sky projection. The UV bright galaxies are seen to be away from the general cluster core, as opposed to the majority of red sequence galaxies.

\section{Morphology}
To better constrain the properties of these UV bright galaxies, we analyse the morphologies of the red sequence galaxies in the sample. We take the two clusters (Abell 2055 and Abell 1650) that overlap area from SDSS (York \etal 2000). With this subsample, we are able to use a variety of different morphological analysis techniques to determine the physical nature of UV bright galaxies and their structural properties.

\begin{figure}
\begin{center}
\includegraphics[scale=0.33]{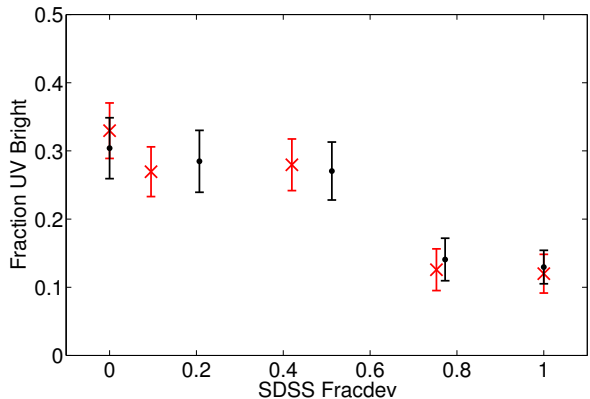}
\caption{Fraction of UV bright galaxies for different SDSS fracDeV R (red crosses) and I (black points) bins. The higher fracDeV galaxies, which assume to have a de Vaucouleurs profile, have less UV bright members compared to the more disk dominated red sequence galaxies.}
\label{fig:fracDeV}
\end{center}
\end{figure}

Using the \textsc{Photoobjall} SDSS dataset, we find all sources corresponding to that of the LARCS red sequence galaxies in both Abell 2055 and 1650, taking the fracDeV parameters from the r and i bands. The fracDeV parameter is a comparative measurement of the surface brightness of a galaxy. A fracDeV of 1 corresponds to a De Vaucouleurs profile, a value of 0 is a pure exponential, with intermediate values a combination of both (see also Masters \etal 2010). Previous studies have used the fracDeV parameter to isolate galaxies, generally believed to be fracDeV $> 0.95$ (e.g. Kaviraj \etal 2007). However, in this study, we can analyse the change in the fracDeV parameter for our red sequence selected galaxies UV properties. Figure \ref{fig:fracDeV} shows the fraction of UV bright galaxies as a function of fracDeV, in all of g and r bands. In this figure the galaxies in the higher fracDeV bins have lower fractions of UV bright galaxies. The UV bright galaxies therefore, appear to be more dominant for galaxies suited to a disk profile, rather than the classical De Vaucouleurs profile.

\begin{figure}
\begin{center}
\includegraphics[scale=0.3]{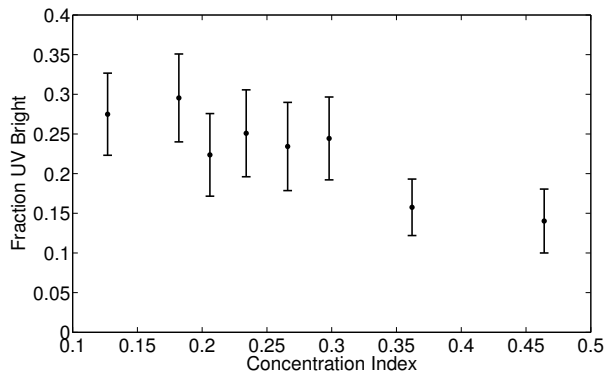}
\caption{Fraction of UV bright galaxies for different concentrations calculated from iraf. The UV bright galaxies appear to be, on average, less prominent in high concentration systems compared to more disky profiles.}
\label{fig:confrac}
\end{center}
\end{figure}

\begin{figure}
\begin{center}
\includegraphics[scale=0.3]{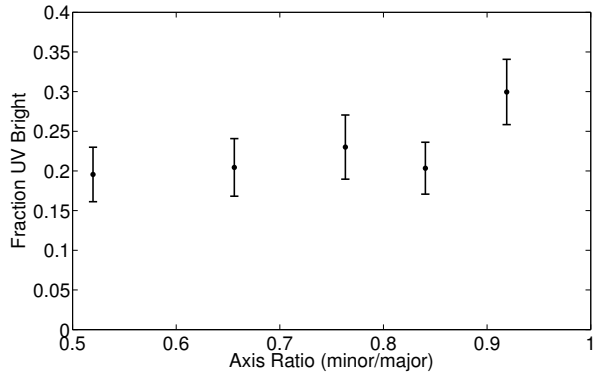}
\caption{UV bright fractions for different axial ratios taken from the SDSS sample. Axial ratios calculated as $A=semiminor/semimajor$ using the iraf command on the LARCS images. The axial ratio appears to have little effect on the UV bright galaxies, suggesting the galaxies are in fact, not predominantly edge on spirals.}
\label{fig:abworld}
\end{center}
\end{figure}

In addition to this, several other photometric parameters from the LARCS dataset are used on the SDSS sample. The central concentration of each galaxy, calculated from the LARCS imaging the R band, and the ratio of semi-minor to semi-major axis are also tested for links to UV bright galaxies. Figures \ref{fig:confrac} and \ref{fig:abworld} both show the UV bright fraction as functions of concentration and semi-minor/semi-major axis respectively. The former figure matches that of figure \ref{fig:fracDeV}, with galaxies having a high concentration containing less UV bright members, compared to the less centrally concentrated, similar to the results from the fracDeV analysis. The axis ratio plot shows an insignificant trend,  with no strong preference to edge on galaxies. The trend from this SDSS sample suggests that UV bright galaxies are less concentrated disk galaxies, compared to the archetypal early-type galaxies. 

To further classify the morphologies within our SDSS sample, we create a sample of red sequence galaxies that overlapped galaxies morphologically classified from Galaxy Zoo\footnote{www.galaxyzoo.org}. In brief, the Galaxy Zoo project is an online project that encourages the general public to vote on the morphological properties of galaxies, and creates probabilities given the relative number of votes (Lintott \etal 2011). Our sample contains only $\sim130$ of the galaxies from the original SDSS sample, a fraction of roughly 20\%. This decreases the robustness of the statistical correction, given the fewer numbers in each bin.

\begin{figure}
        \centering
        \begin{subfigure}[b]{1\textwidth}
                \includegraphics[scale=0.3]{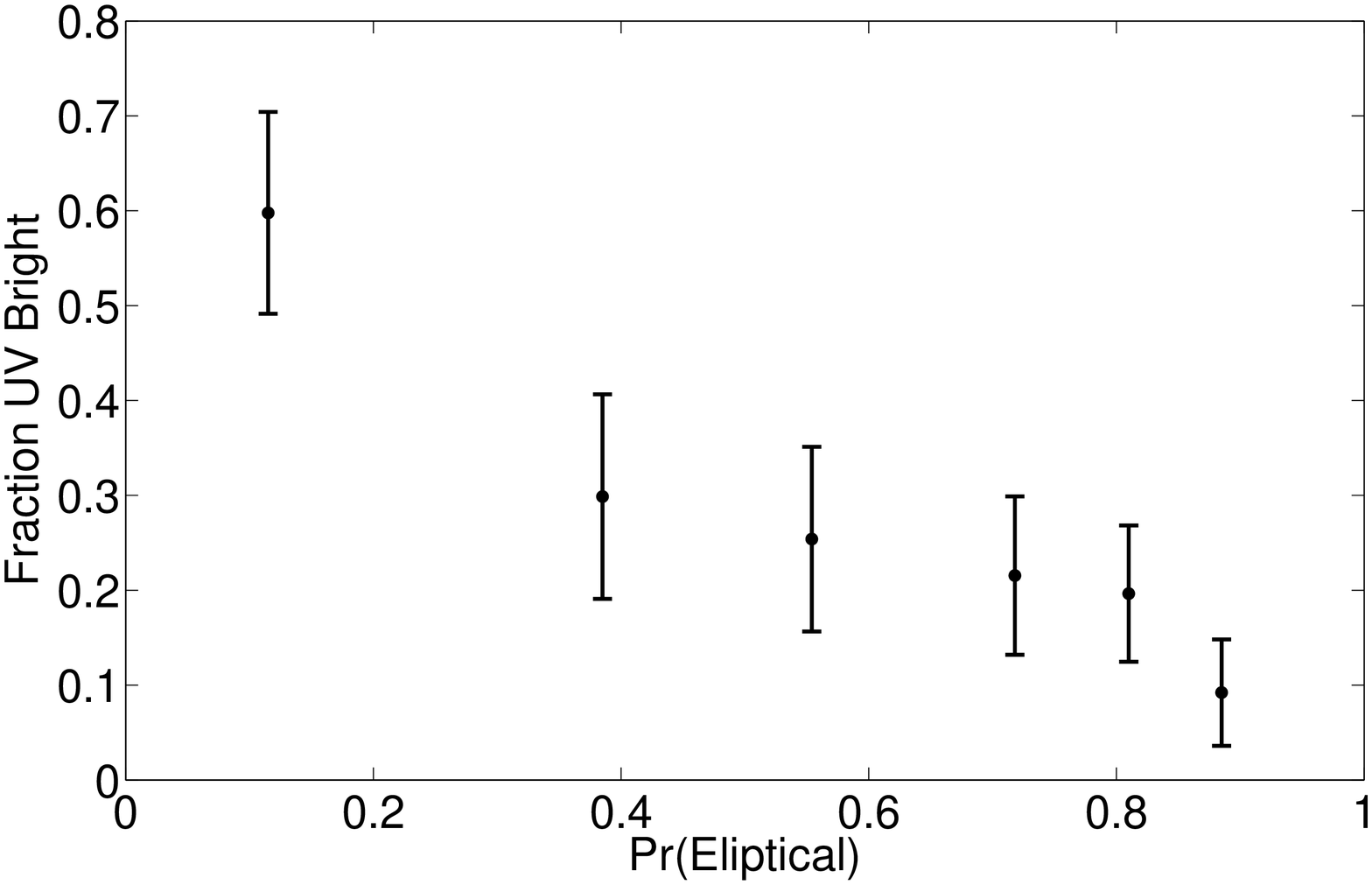}
        \end{subfigure}%
      		\vspace{4mm}
        \begin{subfigure}[b]{1\textwidth}
                \includegraphics[scale=0.3]{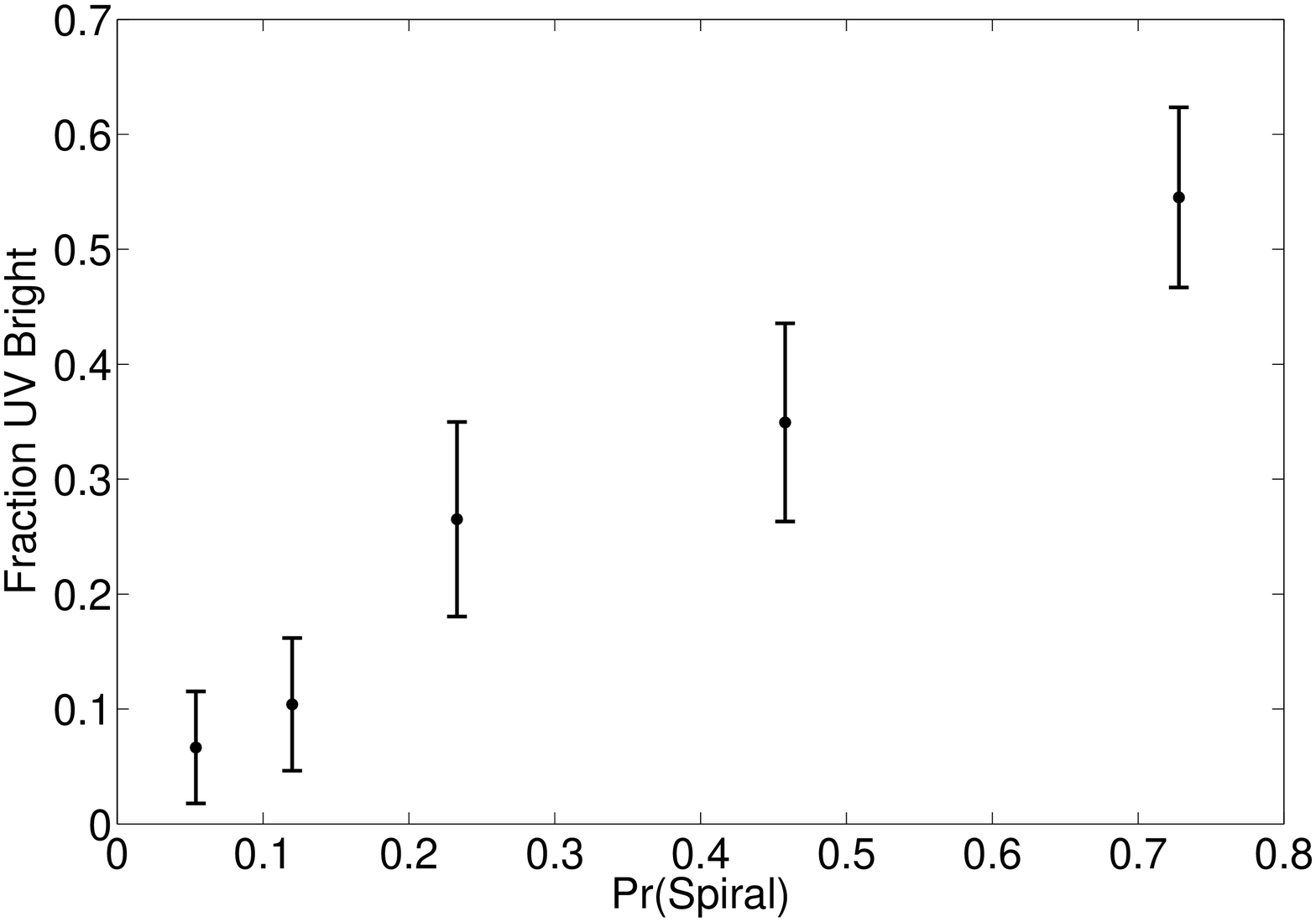}
        \end{subfigure}
       \caption{(top) Fraction of UV bright galaxies  for binned percentage votes of a galaxy being elliptical from the Galaxy Zoo project. Few UV bright galaxies appear to be assigned as elliptical, and instead denoted as spiral (bottom).}
\label{fig:prellip}
\end{figure}

\begin{figure}
\begin{center}
\includegraphics[scale=0.3]{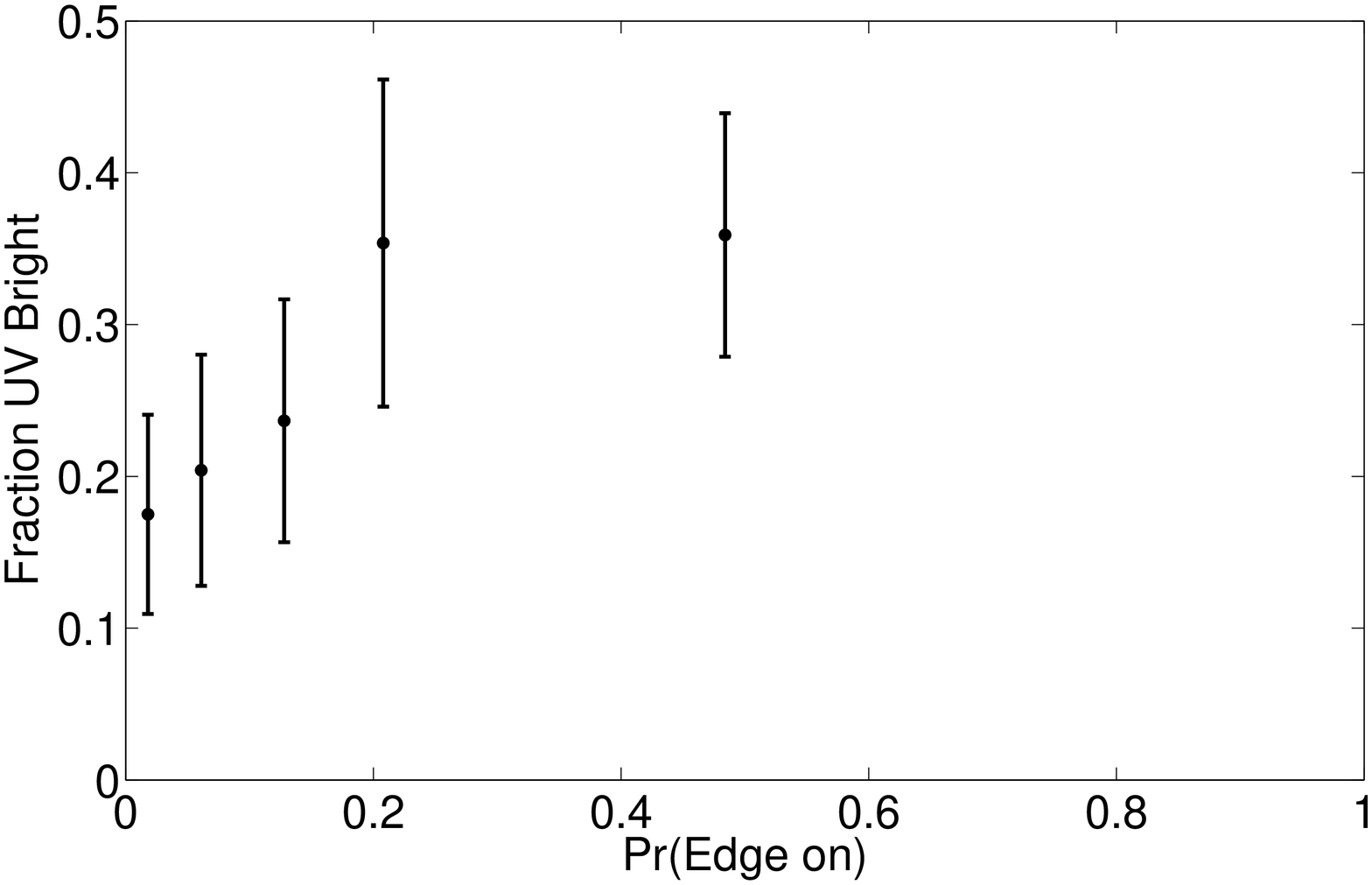}
\caption{UV bright fractions for varied edge on probabilities from Galaxy Zoo. While there seems to be a trend with a galaxy being edge on, it is generally still low, with the mean $pr(edgeon)=0.196$, and few members above Pr(0.5).}
\label{fig:predge}
\end{center}
\end{figure}

Figure \ref{fig:prellip} (top) shows the fraction of UV bright galaxies, for increasing percentage of votes for a galaxy to be ``Elliptical". We see that the fraction of galaxies with high UV decreases with the percentage assigned to the galaxy being a classical elliptical. Similarly, Figure \ref{fig:prellip} (bottom) shows that the UV bright galaxies are more likely to be classified spirals. This is consistent with the larger sample of SDSS galaxies, indicating the UV bright galaxies are in fact, optically red disk galaxies. The chance of edge on galaxies are also tested with Galaxy Zoo, using the P(Edge on) statistic. This figure (\ref{fig:predge}), seems to show a difference in the amount of edge on galaxies than the SDSS sample, with a weak ($<1\sigma$) trend towards UV galaxies being generally edge on. However, all bins contain a median value $<0.5$. The mean of the sample is also low, being $0.2\pm0.2$. While there is an apparent trend, the chance these galaxies are indeed edge on, is still not high from the galaxy zoo data.

\section{Discussion}
In this work we have used NUV optical colour magnitude diagrams to detect the presence of recent star formation in red sequence galaxies. The threshold of the NUV-R=5.4 is adopted, for comparison to other studies (Kaviraj \etal 2007; Schawinski \etal 2007). This threshold adopted is used to disregard NUV contributions from old stellar populations, including low metallicity Horizontal Branch stars, and UV upturn from extremely blue horizontal branch stars. Observations with integral field units should also be able to exclude these galaxies as they will preferentially be slow rotators (Bureau et al.\ 2011). Our results show a trend for these UV bright galaxies to prefer cluster outskirts, with low density environments and disk like profiles. Here we discuss the different scenarios that could cause the blue UV colours of these red sequence galaxies, and the overall implications of such a population.

\subsection{Residual star formation, or red spirals?}
This idea of recent star formation assumes these red sequence galaxies to be morphological early-type galaxies, with either a gas enhancement (Trager \etal 2000) or interactions (Kaviraj \etal 2007; Pierce \etal 2005), causing an extra burst of star formation in otherwise red-and-dead galaxies. The results from Galaxy Zoo and the SDSS morphological analysis of these red sequence galaxies show that UV bright galaxies may not be classical elliptcals as previously expected. The low concentration, disk dominated light curve, and enhanced percentage of being spiral, make these UV bright galaxies seem more like late-type galaxies. This is despite the optical colours of these objects suggesting they are red sequence ellipticals. These galaxies, despite their passive optical colours, show signs of being spiral in morphology, with a small amount of star formation present.

These galaxies do not seem to be edge on spirals that are dust obscured. While the NUV is very sensitive to dust extinction, knots of star formation in the outer disk of a dust obscured spiral could produce this signal a blue NUV colour. The axis ratios of our sample however, show no trend towards UV bright galaxies preferring uneven semimajor and semiminor axes. This is backed up in the Galaxy Zoo sample, with a mean $P(Edge)<0.2$. These galaxies are therefore not preferentially edge on, and while edge on spirals are present, they are not the dominant cause of these UV bright galaxies.

In addition to the lack of asymmetric axial ratios, we use the Galaxy Zoo sample to investigate the fraction of UV bright galaxies undergoing a merger. This is done using the P(Merger) statistic from Galaxy Zoo. For the entire red sequence, an upper limit mean of $P(Merger)=0.011$ is found. This compares to a maximum $P(Merger)=0.010$ for the UV bright sample -- suggesting they are not mergers. We find that 5 out the 131 galaxies have $P(merger)>0.19$, with only 2 being UV bright. While many Galaxy Zoo members have voted for a candidate as a merger conservatively at this redshift, none of our UV bright galaxies are at a percentage of a clear merger of $P(merger)>0.4$ (Darg \etal 2010).  We conclude that these galaxies are therefore not undergoing any mergers, based on low percentage votes in Galaxy Zoo. Additionally, these galaxies do not appear to be fast rotating ellipticals. Our galaxies are seen to prefer low density environments, and have a no preference for an elongated axial ratio. By contrast, fast rotators are ellongated, and prefer higher density regions (Cappellari et al.\ 2011). We argue that these galaxies are more traditional disk-like spirals, and not fast rotator early-type galaxies.

This greatly differs to past results, including that Pierce \etal (2005) and of Kaviraj \etal (2012). These works detailed how merging systems can create a low level of young stars in an older galaxy. Specifically, Kaviraj \etal (2012) demonstrated a link between early-type galaxies with disturbed features and high UV flux. Whilst we have seen a small population of recently star forming elliptical galaxies (cf.\ Yi et al.\ 2005; Salim et al. 2012), the majority of our UV bright sample is dominated by red spiral galaxies. This difference in the present work is probably due to sample selection, where Kaviraj \etal (2012) explicitly selected bulge-like light curves, where this sample only used a colour selection for a red sequence. We don't consider these results to be in contradiction, it is simply the case that a different sample has been employed. Where we have only assumed passive optical colours as a basis for our red sequence, many other works have pre-selected galaxies based on their morphology (e.g. Yi \etal 2005; Kaviraj \etal 2007; Schawinski \etal 2007; Kaviraj \etal 2012), leading to differences in observed results. While it is possible that the morphology criteria are not always accurate (Yi \etal 2005), our UV bright galaxies may represent a different transition state to those presented in other works.

\subsection{Mechanisms causing recent star formation}
If these UV bright spiral galaxies are indeed spirals, their red optical colours may be a signature that these galaxies have had a past truncation of star formation. The morphological and environmental analysis in a cluster setting can help constrain the reason behind these optically red spirals. While the effect of mergers has been discounted, other cosmological effects, particularly from the parent cluster, can be analysed with this sample.

The five clusters used in this sample all have similar fractions of UV bright galaxies, independent of cluster size and regularity. In particular, the cluster Abell 1664, a highly irregular galaxy suggested to be a cluster-cluster merger (Kleiner \etal 2013, in prep), does not show an enhanced amount in UV bright galaxies. The exception to this can be seen in Abell 2055, with a factor of three higher UV bright sources. This is anomalous, given it is an average size, dynamically relaxed cluster compared to the rest of our sample (Pimbblet \etal 2006). This is best explained in the amount of sources detected by \textsl{\textsc{Galex}}
 in this region. While the average number of detected UV sources was around 40,000 to our limiting magnitude, Abell 2055 contained twice as many UV detections, after accounting for a limiting magnitude. We therefore argue that the reason Abell 2055 contains an anomalous amount of UV bright objects is due to cosmic variance. It is apparent from the cluster by cluster analysis that the morphology of the cluster, and the size of the cluster, do not have significant bearing on the stellar populations on the 1 Gyr timescale.

A more common suppression of star formation in clusters is ram pressure stripping. Ram pressure stripping (Gunn and Gott 1972) is caused when galaxies infalling to a large cluster lose their gas content, due to the hot intracluster medium it is moving within. This effect is known to happen often in the cluster core (Quilis 2000), however, can explain decreases in star formation rate at over $2 r_{200}$ (Bahe \etal 2013). This mechanism, while stripping the gas content of a galaxy, can preserve the stellar structure (Gunn \& Gott 1972), and can cause the disk of the galaxy to thicken (Farouki \& Shapiro 1980; Smith et al.\ 2012). The general disk nature, and bulge to disk ratio does not change significantly over larger timescales (Smith et al.\ 2012; Tonnesen \& Bryan 2012). 
This allows the morphology for the galaxy to be preserved, while the gas source for young stars is lost. Our environmental relations suggest that the UV bright galaxies are on the outskirts of the cluster. These galaxies could present evidence of galaxies having experienced a truncation around $\sim 1Gyr$ ago, with the UV signal a relic, or last breath, or star formation from a stripped spiral. The preserved spiral morphology, radial and density relations, as well as the UV bright nature despite red optical colours, present this scenario as a more likely candidate than the others provided.

The alternative explanation to this scenario suggests that these spirals are large in mass, and therefore optically dominated by the older stellar populations. The UV would be therefore picking up star formation in a spiral galaxies, but the optical colours are dominated by the underlying older stars. This would represent galaxies with a normal star formation rate, but with a large mass, would have a lower specific star formation rate. The limitations of the \textsl{\textsc{Galex}}
 AIS survey prevent the use of absolute magnitudes as a proxy for mass in this study, as the amount of faint objects is intrinsically underestimated. This is therefore left as a future endeavour. These two scenarios are lacking in spectral information to both confirm the 3 space location of these galaxies compared to the cluster, and accurate stellar masses. The use of spectral information would best disentangle the star formation rates, and velocity measurements, to determine their formation.

\section{Conclusions}
In this paper, we have quantified the amount of recent star formation in large galaxies clusters from the LARCS survey using \textsl{\textsc{Galex}}
 NUV observations. We have used statistical measurements to get a representative sample of galaxies, with minimal foreground contamination, and tested the properties of UV bright, red sequence galaxies in clusters. Our findings presented are:
\begin{enumerate}
\item Red sequence galaxies with high UV seem to reside on cluster outskirts, with a high radius and low density.
\item These galaxies appear to be generally less concentrated than their passive red sequence counterparts.
\item Despite the passive optical colours, these galaxies appear to generally be morphologically spiral, and not classical bulge dominated systems
\item These systems do not generally appear to be elongated or edge on, and are not generally merging.
\end{enumerate}
These findings have been compiled to constrain the mechanisms possible for a UV bright red sequence galaxy. The most likely scenario is truncation of star formation to a gaseous disk galaxy through a stripping and heating of available gas, denying the formation of new stars as the galaxies move into the cluster. The UV observed is due to the last remaining gas in the galaxies forming stars, a last gasp of star formation as it moves into the cluster centre. This transition state will only exist for a short time, before these galaxies both lose their last signals of star formation. Alternatively, these galaxies may be high-mass spiral galaxies that are dominated by an older stellar population, with normal star formation still occurring (i.e. a normal star formation rate, but a low specific star formation rate). In either sense, these disk galaxies demonstrate a shortcoming of the optical colour-magnitude relation, but shows the UV as a tracer for low specific star formation rate galaxies. This can be used to isolate red spiral galaxies with a predominant old stellar population base, whether they be due to truncation, or high mass.

\section*{Acknowledgements}
We thank the anonymous referee for her/his feedback which has improved the work presented in this manuscript.
JPC acknowledges financial support through the Monash Dean's Scholarship.  
KAP thanks Christ Church College, Oxford, for their hospitality whilst some of this work was being undertaken

Funding for the SDSS and SDSS-II has been provided by the Alfred P. Sloan Foundation, the Participating Institutions, the National Science Foundation, the U.S. Department of Energy, the National Aeronautics and Space Administration, the Japanese Monbukagakusho, the Max Planck Society, and the Higher Education Funding Council for England. The SDSS Web Site is http://www.sdss.org/.

The SDSS is managed by the Astrophysical Research Consortium for the Participating Institutions. The Participating Institutions are the American Museum of Natural History, Astrophysical Institute Potsdam, University of Basel, University of Cambridge, Case Western Reserve University, University of Chicago, Drexel University, Fermilab, the Institute for Advanced Study, the Japan Participation Group, Johns Hopkins University, the Joint Institute for Nuclear Astrophysics, the Kavli Institute for Particle Astrophysics and Cosmology, the Korean Scientist Group, the Chinese Academy of Sciences (LAMOST), Los Alamos National Laboratory, the Max-Planck-Institute for Astronomy (MPIA), the Max-Planck-Institute for Astrophysics (MPA), New Mexico State University, Ohio State University, University of Pittsburgh, University of Portsmouth, Princeton University, the United States Naval Observatory, and the University of Washington.

This work has made use of public \textsl{\textsc{Galex}} data. \textsl{\textsc{Galex}} is a NASA Small Explorer, launched in 2003 April. We gratefully acknowledge NASA's support for construction, operation and science analysis for the \textsl{\textsc{Galex}} mission, developed in cooperation with the Centre National d'Etudes Spatiales of France and the Korean Ministry of Science and Technology.


\begin{thebibliography}{}

\bibitem[\protect\citeauthoryear{Bahe et al.}{2013}]{2013MNRAS...430...3017}
Bah\'{e} Y. M., McCarthy I. G., Balogh, M. L., Font, A. S., 2013, MNRAS, 430, 3017

\bibitem[\protect\citeauthoryear{Bower et al.}{1992}]{1992MNRAS...254...601}
Bower, R., G., Lucey, J., R., \& Ellis, R., S., 1992, MNRAS, 254, 601

\bibitem[\protect\citeauthoryear{Bureau et al.}{2011}]{2011MNRAS.414.1887B}
Bureau, M., et al., 2011, MNRAS, 414, 1887

\bibitem[\protect\citeauthoryear{Burstein et al.}{1988}]{1988ApJ...328...440}
Burstein, D., Gertola, F., Buson, L., M., Faber, S., M., \& Lauer, T., R., 1988, ApJ, 328, 440

\bibitem[\protect\citeauthoryear{Cappelari et al.}{2011}]{2011MNRAS...416...1680}
Cappellari, M., et al., 2011, MNRAS, 416, 1680

\bibitem[\protect\citeauthoryear{Carlberg, Yee, \& Ellingson}{1997}]{1997ApJ...478..462C} 
Carlberg R.~G., Yee H.~K.~C., Ellingson E., 1997, ApJ, 478, 462 

\bibitem[\protect\citeauthoryear{CodeWelch}{1979}]{1979ApJ...228...95C}
Code, A., D., \& Welch, G., A., 1979, ApJ, 228, 95

\bibitem[\protect\citeauthoryear{Cotese}{2013}]{2013A&A...543...132}
Cortese, L., 2013, A\&A, 543, 132 

\bibitem[\protect\citeauthoryear{Darg et. al.}{2010}]{2010MNRAS...401...1043}
Darg, D, W., et. al., 2010, MNRAS, 401, 1043

\bibitem[\protect\citeauthoryear{Dressler}{1980}]{1980ApJ...236...351}
Dressler, A., 1980, ApJ, 236, 351

\bibitem[\protect\citeauthoryear{Drinkwater}{2010}]{2010MNRAS...401...1429}
Drinkwater, M. J., et al., 2010, MNRAS, 401, 1429

\bibitem[\protect\citeauthoryear{Faber}{1973}]{1973ApJ...179..731F}
Faber, S. M., 1973, ApJ, 179, 731

\bibitem[\protect\citeauthoryear{Farouki \& Shapiro}{1980}]{1980ApJ...241...928}
Farouki R., \& Shapiro S. L., 1980, ApJ, 241, 928

\bibitem[\protect\citeauthoryear{Ferreras \& Silk}{2000}]{2000ApJL...541...L37}
Ferreras, I., \& Silk, J., 2000, ApJL, 541, L37

\bibitem[\protect\citeauthoryear{Greggio \& Renzini}{1990}]{1990ApJ...364...35}
Greggio, L. \& Renzini A., 1990, ApJ, 364, 35

\bibitem[\protect\citeauthoryear{Gunn \& Gott}{1972}]{1972ApJ...176...1}
Gunn, J. E., \& Gott J. R., 1972, ApJ, 176, 1

\bibitem[\protect\citeauthoryear{Han et al.}{2007}]{2007MNRAS...380...1098}
Han, Z., Podsiadlowski, Ph., \& Lynas-Gray, A. E. 2007, MNRAS,
380, 1098

\bibitem[\protect\citeauthoryear{Huertas-Company et al.}{2010}]{2010AA...515...3}
Huertas-Company, M., Aguerri, J. A. L., Tresse, L., Bolzonella, M., Koekemoer, A. M., \& Maier, C., 2010, A\&A, 515, 3

\bibitem[\protect\citeauthoryear{Jeong et al.}{2009}]{2009MNRAS...398...2028}
Jeong, H., et al. 2009, MNRAS, 398, 2028

\bibitem[\protect\citeauthoryear{Kaviraj et al.}{2007}]{2007ApJS...173...619}
Kaviraj, S., et al. 2007, ApJS, 173, 619

\bibitem[\protect\citeauthoryear{Kaviraj, Tan, Ellis, \& Silk}{2011}]{2011MNRAS...411...2148}
Kaviraj, S., Tan, K,-M., Ellis, R, S., \& Silk, J., 2011, MNRAS, 411, 2148

\bibitem[\protect\citeauthoryear{Kodama et al.}{1998}]{1998A&A...334...99}
Kodama, T., Arimoto, N., Barger, A.J., \& Arag'on-Salamanca, A., 1998, A\&A 334, 99

\bibitem[\protect\citeauthoryear{Kohno et al.}{2002}]{2002PASJ...54...541}
Kohno, K., Tosaki, T., Matsushita, S., Vila-Vila\'{o}, B., Shibatsuka, T., Kawabe, R., 2002 PASJ, 54, 541

\bibitem[\protect\citeauthoryear{Landolt}{1992}]{1992AJ...104...340}
Landolt A. U., 1992, AJ, 104, 340

\bibitem[\protect\citeauthoryear{Lintott et al.}{2011}]{2011MNRAS.410..166}
Lintott C. J. et al., 2011, MNRAS, 410, 166

\bibitem[\protect\citeauthoryear{MacLaren, Ellis \& Couch}{1987}]{1987MNRAS...230...249}
MacLaren, I., Ellis, R. S., Couch, W. J., 1988, MNRAS, 230, 249

\bibitem[\protect\citeauthoryear{McIntosh et al.,}{2013}]{2013MNRAS}
McIntosh, D. H., et al., 2013, MNRAS, submitted (arXiv:1308.0054)

\bibitem[\protect\citeauthoryear{Martin et al.}{2005}]{2005ApJ...619L...1M} 
Martin D.~C., et al., 2005, ApJ, 619, L1 

\bibitem[\protect\citeauthoryear{Masters et al.}{2010}]{2010MNRAS...405...783} 
Masters, K., L., et al., 2010, MNRAS, 405, 783 

\bibitem[\protect\citeauthoryear{Morrissey et al.}{2007}]{2007ApJS...173...682}
Morrissey, P., et al., 2007, ApJS, 173, 682

\bibitem[\protect\citeauthoryear{Muldrew et al.}{2012}]{2012MNRAS...419...2670}
Muldrew, S., I.,  et al., 2012, MNRAS, 419, 2670

\bibitem[\protect\citeauthoryear{Oconnerll}{1999}]{1999ARA&A...37...603}
O'Connell R. W., 1999, ARA\&A, 37, 603

\bibitem[\protect\citeauthoryear{Peng \& Nagai}{2009}]{2009ApJ...705...58}
Peng, F., \& Nagai, D. 2009, ApJ, 705, 58

\bibitem[\protect\citeauthoryear{Pierce et al.}{2005}]{2005MNRAS...358...419}
Pierce, M., Brodie, J., P., Forbes, D., A., Beasley, M., A., Proctor, R., \& Strader, J., 2005, MNRAS, 358, 419

\bibitem[\protect\citeauthoryear{Pimbblet et 
al.}{2001}]{2001MNRAS.327..588P} Pimbblet K.~A., Smail I., Edge A.~C., Couch W.~J., O'Hely E., Zabludoff A.~I., 2001, MNRAS, 327, 588 

\bibitem[\protect\citeauthoryear{Pimbblet et 
al.}{2002}]{2002MNRAS.331..333P} Pimbblet K.~A., Smail I., Kodama T., Couch W.~J., Edge A.~C., Zabludoff A.~I., O'Hely E., 2002, MNRAS, 331, 333 

\bibitem[\protect\citeauthoryear{Pimbblet et 
al.}{2006}]{2006MNRAS.366..645P} Pimbblet K.~A., Smail I., Edge A.~C., O'Hely E., Couch W.~J., Zabludoff A.~I., 2006, MNRAS, 366, 645 

\bibitem[\protect\citeauthoryear{Pimbblet \& Couch}{2012}]{2012MNRAS...419...1153}
Pimbblet, K. A., \& Couch, W. J., 2012, MNRAS, 419, 1153

\bibitem[\protect\citeauthoryear{Quilis, V., Moore, B., \& Bower}{2000}]{2000Science...288...1617}
Quilis, V., Moore, B., \& Bower, R., 2000, Science, 288, 1617

\bibitem[\protect\citeauthoryear{Rawle et al.}{2008}]{2008}
Rawle, Timothy D.; Smith, Russell J.; Lucey, John R.; Hudson, Michael J.; Wegner, Gary A., 2008, MNRAS, 385, 2097

\bibitem[\protect\citeauthoryear{Sadat et al.}{2004}]{2004A&A...424.1097S} Sadat R., 
Blanchard A., Kneib J.-P., Mathez G., Madore B., Mazzarella J.~M., 2004, A\&A, 424, 1097

\bibitem[\protect\citeauthoryear{Salim et al.}{2012}]{2012ApJ...755..105}
Salim, S., Fang, J. J., Rich, R. M., Faber, S. M., \& Thilker, D. A., 2012, ApJ, 755, 105

\bibitem[\protect\citeauthoryear{S{\'a}nchez-Bl{\'a}zquez et 
al.}{2009}]{2009MNRAS.400.1264S} S{\'a}nchez-Bl{\'a}zquez P., Gibson B.~K., Kawata D., Cardiel N., Balcells M., 2009, MNRAS, 400, 1264 

\bibitem[\protect\citeauthoryear{Schawinski et al.}{2007}]{2007ApJ...173...512}
Schawinski, K., et al., 2007, ApJ, 173, 512

\bibitem[\protect\citeauthoryear{Schawinski et al.}{2009}]{2009MNRAS.396..818S2}
Schawinski, K., et al., 2009, MNRAS, 396, 818

\bibitem[\protect\citeauthoryear{Serra \& Trager}{2006}]{2006MNRAS...374...769}
Serra, P., \& Trager, S., C., 2007, MNRAS, 374, 769

\bibitem[\protect\citeauthoryear{Shapiro et al.}{2010}]{2010MNRAS.402.2140}
Shapiro, K. L., et al., 2010, MNRAS, 402, 2140

\bibitem[\protect\citeauthoryear{Smith et al.}{2012}]{2012MNRAS...420...1990}
Smith, R. Fellhauer, M., \& Assmann, P., 2012, MNRAS, 420, 1990

\bibitem[\protect\citeauthoryear{Smith et al.}{2011}]{2011MNRAS...421...2982}
Smith, R. J., Lucey, J. R., \& Carter, D., 2011, MNRAS, 421, 2982

\bibitem[\protect\citeauthoryear{Tonnesen \& Bryan}{2012}]{2009MNRAS.394.2098S} 
Tonnesen, S., \& Bryan, G., L., .2012, MNRAS, 422, 1609

\bibitem[\protect\citeauthoryear{Tojeiro et al.}{2013}]{2013MNRAS.432..359}
Tojeiro, R., et al., 2013, MNRAS, 432, 359

\bibitem[\protect\citeauthoryear{Tosaki \& Shioya}{1997}]{1997ApJ....484..664} 
 Tosaki, T., \& Shioya, Y., 1997 ApJ 484, 664

\bibitem[\protect\citeauthoryear{Trager et al.}{2000}]{2000AJ....120..165T} 
Trager S.~C., Faber S.~M., Worthey G., Gonz{\'a}lez J.~J., 2000, AJ, 120, 165 

\bibitem[\protect\citeauthoryear{Urquhart, Willis, Hoekstra, Pierre}{2005}]{2010MNRAS...406...368}
Urquhart S. A., Willis J. P., Hoekstra H., Pierre M., 2010, MNRAS, 406, 368

\bibitem[\protect\citeauthoryear{Valentinuzzi et al.}{2011}]{2011A&A...536...17}
Valentinuzzi, T., et al.\ 2011, A\&A, 536, 17

\bibitem[\protect\citeauthoryear{Visvanthan \& Sandage}{1977}]{1977ApJ...216...214}
Visvanathan, N., \& Sandage, A., 1977, ApJ, 216, 214

\bibitem[\protect\citeauthoryear{Wolf et al.}{2009}]{2009MNRAS...393...1302}
Wolf C. et al., 2009, MNRAS, 393, 1302

\bibitem[\protect\citeauthoryear{Yi et al.}{1997}]{1997ApJ...482..677Y}
Yi, S. K., Demarque, P., Oemler, A., , 1997, ApJ, 482, 677

\bibitem[\protect\citeauthoryear{Yi et al.}{2005}]{2005ApJL...619...L111}
Yi, S., K., et al. 2005, ApJL, 619, L111

\bibitem[\protect\citeauthoryear{Yi et al.}{2011}]{2011ApJS...195...22}
Yi, S. K., Lee, J., Sheen, Y., Jeong, H., Suh, H., \& Oh, K., 2011, ApJS 195, 22

\bibitem[\protect\citeauthoryear{York et al.}{2000}]{2000AJ...120...1579}
York, D.G., Adelman, J., Anderson, J.E., et al. 2000, AJ, 120, 1579


\end{thebibliography}
\end{document}